\documentclass[letterpaper, 12pt]{article}[2000/05/19]
\usepackage[english]{babel}
\usepackage{amsfonts,amsmath,amssymb,amsthm,latexsym,amscd,mathrsfs}
\usepackage{ifthen,cite}
\usepackage[bookmarksnumbered=true]{hyperref}

\hypersetup{pdfpagetransition={Split}}

\newcommand{\evenhead}{Author \ name}
\newcommand{\oddhead}{Article \ name}
\newcommand{\theArticleName}{Article name}

\newcommand{\FirstPageHeading}[1]{\thispagestyle{empty}%
\noindent\raisebox{0pt}[0pt][0pt]{\makebox[\textwidth]{\protect\footnotesize \sf }}\par}

\newcommand{\ArticleName}[1]{\renewcommand{\theArticleName}{#1}\vspace{-2mm}\par\noindent {\LARGE\bf  #1\par}}
\newcommand{\Author}[1]{\vspace{5mm}\par\noindent {\Large  #1\par} \par\vspace{2mm}\par}
\newcommand{\Address}[1]{\vspace{2mm}\par\noindent {\it #1} \par}
\newcommand{\Email}[1]{\ifthenelse{\equal{#1}{}}{}{\par\noindent {\rm E-mail: }{\it  #1} \par}}
\newcommand{\URLaddress}[1]{\ifthenelse{\equal{#1}{}}{}{\par\noindent {\rm URL: }{\tt  #1} \par}}
\newcommand{\EmailD}[1]{\ifthenelse{\equal{#1}{}}{}{\par\noindent {$\phantom{\dag}$~\rm E-mail: }{\it  #1} \par}}
\newcommand{\URLaddressD}[1]{\ifthenelse{\equal{#1}{}}{}{\par\noindent {$\phantom{\dag}$~\rm URL: }{\tt  #1} \par}}

\newcommand{\Abstract}[1]{\vspace{6mm}\par\noindent\hspace*{10mm}
\parbox{140mm}{\small {\bf Abstract.} #1}\par}
\newcommand{\Keywords}[1]{\vspace{3mm}\par\noindent\hspace*{10mm}
\parbox{140mm}{\small {\bf Key words:} \rm #1}\par}
\newcommand{\Classification}[1]{\vspace{3mm}\par\noindent\hspace*{10mm}
\parbox{140mm}{\small {\it 2000 Mathematics Subject Classification:} \rm #1}\vspace{3mm}\par}
\newcommand{\ShortArticleName}[1]{\renewcommand{\oddhead}{#1}}
\newcommand{\AuthorNameForHeading}[1]{\renewcommand{\evenhead}{#1}}

\long\def\@makecaption#1#2{
  \sbox\@tempboxa{\small \textbf{#1.}\ \ #2}%
  \ifdim \wd\@tempboxa >\hsize
    {\small \textbf{#1.}\ \ #2}\par \else
    \global \@minipagefalse
    \hb@xt@\hsize{\hfil\box\@tempboxa\hfil}%
  \fi \vskip\belowcaptionskip}

\def\numberwithin#1#2{\@ifundefined{c@#1}{\@nocounterr{#1}}{%
  \@ifundefined{c@#2}{\@nocnterr{#2}}{%
  \@addtoreset{#1}{#2}%
  \toks@\@xp\@xp\@xp{\csname the#1\endcsname}%
  \@xp\xdef\csname the#1\endcsname
    {\@xp\@nx\csname the#2\endcsname.\the\toks@}}}}
\def\E^#1{{\buildrel #1 \over\vee}}

{\theoremstyle{definition}

}

\begin{document}

\FirstPageHeading{Yu.Yu. Fedchun and V.I. Gerasimenko}

\ShortArticleName{Evolution equations in functional derivatives}

\AuthorNameForHeading{Yu.Yu. Fedchun and V.I. Gerasimenko}

\ArticleName{Evolution Equations in Functional Derivatives\\ of
Many-Particle Systems}
\Author{Yu.Yu. Fedchun$^\ast$\footnote{E-mail:\emph{fedchun.yu@ukr.net}} and
        V.I. Gerasimenko$^\ast$$^\ast$\footnote{E-mail:\emph{gerasym@imath.kiev.ua}}}

\Address{$^\ast$\hspace*{1mm}Taras Shevchenko National University of Kyiv,\\
    \hspace*{4mm}Department of Mechanics and Mathematics,\\
    \hspace*{4mm}2, Academician Glushkov Av.,\\
    \hspace*{4mm}03187 Kyiv, Ukraine}

\Address{$^\ast$$^\ast$Institute of Mathematics of NAS of Ukraine,\\
         \hspace*{3mm}3, Tereshchenkivs'ka Str.,\\
         \hspace*{3mm}01601 Kyiv-4, Ukraine}
\bigskip

\Abstract{The hierarchies of evolution equations of classical many-particle systems are formulated
as evolution equations in functional derivatives. In particular the BBGKY hierarchy for marginal
distribution functions, the dual BBGKY hierarchy for marginal observables, the Liouville hierarchy
for correlation functions and the nonlinear BBGKY hierarchy for the marginal correlation functions
are considered. The nonperturbative solution expansions of the Cauchy problem of these hierarchies
are constructed on the basis of established relations between the generating functionals of corresponding
functions. The obtained results are generalized on systems of particles interacting via many-body potentials.}

\bigskip

\vspace{2pc}
\Classification{82C05; 82C22; 49S05; 47J30; 35Q82.}

\makeatletter
\renewcommand{\@evenhead}{
\hspace*{-3pt}\raisebox{-15pt}[\headheight][0pt]{\vbox{\hbox to
\textwidth {\thepage \hfil \evenhead}\vskip4pt \hrule}}}
\renewcommand{\@oddhead}{
\hspace*{-3pt}\raisebox{-15pt}[\headheight][0pt]{\vbox{\hbox to
\textwidth {\oddhead \hfil \thepage}\vskip4pt\hrule}}}
\renewcommand{\@evenfoot}{}
\renewcommand{\@oddfoot}{}
\makeatother

\newpage
\vphantom{math} \protect\tableofcontents

\vspace{0.5cm}

\section{Introduction}

It is well known the meaning of methods of the functional differentiation and functional integration
for the problems of mathematical physics \cite{V,BSh} and applications \cite{DE}, in particular with
regard to many-particle systems of statistical mechanics or quantum field theory. As far as we know
for the first time equations in functional derivatives for the description of the evolution of states
of many-particle systems were used by T.H. Gronwall \cite{G15}, M.A. Leontovich \cite{Leon} and N.N.
Bogolyubov \cite{B1,B2}. Bogolyubov's results were developed by R.L. Lewis in the paper \cite{L}, where
it was constructed the reduced representation of a nonperturbative solution of the Cauchy problem of the
BBGKY hierarchy. On quantum many-particle systems Bogolyubov approach was extended in \cite{AP}.

In this article we present the rigorous results on the evolution equations in functional derivatives
for generating functionals of states and observables of classical many-particle systems, namely the
BBGKY hierarchy and the dual BBGKY hierarchy in functional derivatives respectively \cite{GeF}. One
more approach to the description of the evolution of states of many-particle systems is given by means
of generating functionals for correlation functions. The generating functionals of correlation functions
and marginal correlation functions are governed by the von Neumann hierarchy and the nonlinear BBGKY
hierarchy in functional derivatives  respectively. On the basis of the developed approach we construct
nonperturbative solutions of the Cauchy problem of corresponding evolution equations in appropriate
Banach spaces.

We briefly outline the results and structure of the paper.
In section 2 we set forth an approach to the description of the evolution of states of many-particle
systems within the framework of the generating functional of marginal distribution functions governed
by the BBGKY hierarchy in functional derivatives. Then by means of introduced generating functional of
marginal distribution functions we construct a nonperturbative solution of the initial-value problem
of the BBGKY hierarchy on the basis of cluster expansions of groups of operators which describe dynamics
of finitely many particles.

In section 3 we develop an alternative method to the description of the evolution of states within the
framework of the generating functionals of correlation functions and marginal correlation functions
governed by the von Neumann hierarchy and the nonlinear BBGKY hierarchy in functional derivatives
respectively. The relationships of the introduced generating functionals of states are established.

In section 4 we state an equivalent approach to the description of the evolution of many-particle
systems within the framework of the generating functional of marginal observables governed by the dual
BBGKY hierarchy in functional derivatives. A nonperturbative solution of the initial-value problem
of the dual BBGKY hierarchy is constructed on the basis of cluster expansions of groups of operators
which describe the evolution of observables.

Finally in section 5 we conclude with some observations and perspectives for future research.


\section{The BBGKY hierarchy in functional derivatives}

\subsection{A generating functional of marginal distribution functions}
Let $f=(f_0,f_1(x_1),\ldots,f_n(x_1,\ldots,x_n),\ldots)$ be a sequence of measurable functions
defined on the phase spaces of the corresponding number of particles and $u=(1,u(x_1),\ldots,
\prod_{i=1}^nu(x_i),\ldots)$ is a sequence of products of continuously differentiable integrable
functions $u(x_i),\,i\geq1$. The generating functional of functions $f_n,\,i\geq1$ is defined by
the following expression
\begin{eqnarray}\label{def_fun}
  &&(f,u)\doteq\sum_{n=0}^\infty\frac{1}{n!}\int\ldots\int
    f_n(x_1,\ldots,x_n)\prod_{i=1}^n u(x_i)dx_1\ldots dx_n.
\end{eqnarray}
For the integrable functions $u(x_i),\,i\geq1$, functional \eqref{def_fun} exists. We refer to
functional (\ref{def_fun}) as the generating functional of functions $f_n(x_1,\ldots,x_n),\,n\geq1$,
by reason of the equality
\begin{eqnarray}\label{def_funder}
  &&f_n(x_1,\ldots,x_n)=\frac{\delta^n}{\delta u(x_1)\ldots\delta u(x_n)}(f,u)\mid_{u=0},
\end{eqnarray}
where $\delta^n/\delta u(x_1)\ldots\delta u(x_n)$ is the $nth$ order functional derivative
(the $nth$ order G\^{a}teaux derivative \cite{GF,FSG}).

We consider a system of a non-fixed number of particles in the space $\mathbb{R}^3$ defined
by the Hamiltonian \cite{CGP97}
\begin{eqnarray}\label{ham1}
  &&H_n=\sum_{i=1}^n K(p_i)+\sum_{i<j=1}^n\Phi(q_i-q_j),
\end{eqnarray}
where $K(p_i)$ is a kinetic energy of the $i$ particle and $\Phi(q_i-q_j)$ is a two-body
interaction potential. Hereafter we assume that the function $\Phi$ satisfies conditions
which guarantee the existence of a global in time solution of the Hamilton equations for
finitely many particles \cite{CGP97}.
The evolution of states of a system is described by the sequence $D(t)=(1,D_1(t,x_1),
\dots,D_n(t,x_1,\ldots,x_n),\ldots)$ of distribution functions. The integrable function
$D_n(t,x_1,\ldots,x_n)$ is defined on the $n$-particle phase space and symmetric with respect
to the permutations of the arguments $x_1,\ldots,x_s$. We introduce the generating functional
$(D(t),u)$ of the distribution functions $D_n(t,x_1,\ldots,x_n),\,n\geq1$, according to
definition (\ref{def_fun}) by the expansion
\begin{eqnarray}\label{func_stan}
  &&(D(t),u)\doteq\sum_{n=0}^\infty\frac{1}{n!}\int\ldots\int
     D_n(t,x_1,\ldots,x_n)\prod_{i=1}^nu(x_i)dx_1\ldots dx_n.
\end{eqnarray}

The evolution of states of many-particle systems in terms of generating functional
(\ref{func_stan}) is governed by the following evolution equation in functional derivatives
\begin{eqnarray}\label{func_liouv}
  &&\frac{d}{dt}(D(t),u)=
    \int \big\{K(p_1),\frac{\delta}{\delta u(x_1)}(D(t),u)\big\}u(x_1)dx_1+\\
  &&+\frac{1}{2!}\int\int\big\{\Phi(q_1-q_2),\frac{\delta^2}
    {\delta u(x_1)\delta u(x_2)}(D(t),u)\big\}u(x_1)u(x_2)dx_1dx_2,\nonumber
\end{eqnarray}
where $\{\,.\,,\,.\,\}$ is the Poisson bracket. Then taking into account equality (\ref{def_funder}),
we observe that the distribution functions $D_n(t),\,n\geq1$, are determined by a sequence of the
Liouville equations
\begin{eqnarray}\label{rivLiyv}
   &&\frac{\partial}{\partial{t}}D_n(t)=\{H_n,D_n(t)\}, \quad n\geq1 .
\end{eqnarray}
We note that the inverse proposition is also true, namely the generating functional of solutions
of the Liouville equations \eqref{rivLiyv} satisfies equation \eqref{func_liouv}. By reason of this
we are referred to evolution equation \eqref{func_liouv} as the Liouville equation in functional
derivatives.

We consider an equivalent approach to the description of the evolution of states of many-particle
systems within the framework of a generating functional of marginal distribution functions
\cite{CGP97}. To that end we introduce a generating functional of the marginal distribution
functions $F_s(t,x_1,\ldots,x_s),\,s\geq1$, by means of generating functional \eqref{func_stan}
\begin{eqnarray}\label{FuconDu}
   &&(F(t),u)=(\widetilde{D}(t),u+1),
\end{eqnarray}
where $\widetilde{D}(t)=(D(t),I)^{-1}(1,D_1(t,x_1),\ldots,D_n(t,x_1,\ldots,x_n),\ldots)$ is a sequence
of normalized distribution functions, $(D(t),I)$ is a normalizing factor (grand canonical partition
function)
\begin{eqnarray*}
   &&(D(t),I)\doteq\sum_{n=0}^\infty\frac{1}{n!}\int\ldots\int D_n(t,x_1,\ldots,x_n)dx_1\ldots dx_n,
\end{eqnarray*}
and hence $I\equiv(1,\ldots,1,\ldots)$. In consequence of introduced generating functional
\eqref{FuconDu} we establish the relation between marginal distribution functions and distribution
functions. For this purpose we transform the generating functional $(\widetilde{D}(t),u+1)$ to form
\eqref{def_fun} to determine the sequence of functions generated by such functional. The following
equality holds
\begin{eqnarray}\label{r1}
  &&(\widetilde{D}(t),u+1)=(e^\mathfrak{a}\widetilde{D}(t),u),
\end{eqnarray}
where on $f\in L^{1}$ the operator $\mathfrak{a}$ (an analog of the annihilation operator \cite{CGP97})
is defined by the formula
\begin{eqnarray}\label{opann}
    &&(\mathfrak{a}f)_n(x_{1},\ldots,x_{n})\doteq\int f_{n+1}(x_{1},\ldots,x_{n+1})dx_{n+1}.
\end{eqnarray}
Indeed we have
\begin{eqnarray*}
  &&(\widetilde{D}(t),u+1)=(D(t),I)^{-1}\sum_{n=0}^\infty\frac{1}{n!}\int\ldots\int
     D_n(t,x_{1},\ldots,x_{n})\prod_{i=1}^n(u(x_i)+1)dx_1\ldots dx_n=\\
  &&=(D(t),I)^{-1}\sum_{n=0}^{\infty}\frac{1}{n!}\int\ldots\int D_n(t,x_{1},\ldots,x_{n})
     \sum_{k=0}^n\sum_{i_1<\ldots<i_k=1}^n u(x_{i_1})\ldots u(x_{i_k})dx_1\ldots dx_n=\\
  &&=(D(t),I)^{-1}\sum_{s=0}^{\infty}\frac{1}{s!}\sum_{n=0}^{\infty}\frac{1}{n!}\int\ldots\int
     D_{s+n}(t,x_{1},\ldots,x_{s+n})\prod_{i=1}^s u(x_i)dx_1\ldots dx_{s+n}.
\end{eqnarray*}
Therefore, according to \eqref{def_funder}, from \eqref{FuconDu} and \eqref{r1} we obtain the expression for
marginal distribution functions within the framework of nonequilibrium grand canonical ensemble \cite{CGP97}
\begin{eqnarray*}
 &&F_s(t,x_1,\ldots,x_s)=(D(t),I)^{-1}\sum_{n=0}^\infty\frac{1}{n!}\int\ldots\int
     D_{s+n}(t,x_{1},\ldots,x_{s+n})dx_{s+1}\ldots dx_{s+n}.
\end{eqnarray*}

Using the Liouville equation in functional derivatives (\ref{func_liouv}), from relation \eqref{FuconDu}
we derive the evolution equation in functional derivatives for the generating functional of marginal
distribution functions
\begin{eqnarray}\label{fh1}
  &&\frac{d}{dt}(F(t),u)=
    \int\big\{K(p_1),\frac{\delta}{\delta u(x_1)}(F(t),u)\big\}(u(x_1)+1)dx_1+\\
  &&+\frac{1}{2!}\int\int\big\{\Phi(q_1-q_2),\frac{\delta^2}
    {\delta u(x_1)\delta u(x_2)}(F(t),u)\big\}\prod_{i=1}^2(u(x_i)+1)dx_1dx_2.\nonumber
\end{eqnarray}
We are referred to this equation as the BBGKY hierarchy in functional derivatives \cite{B1,B2,L}.
Taking into account equality (\ref{def_funder}), from evolution equation \eqref{fh1} we establish
that the marginal distribution functions $F_s(t),\,s\geq1$, are determined by the BBGKY hierarchy
\begin{eqnarray}\label{hierarBBGKI}
   &&\frac{\partial}{\partial t}F_{s}(t)=\big\{H_s,F_{s}(t)\big\}+
      \int\big\{\sum\limits_{i=1}^{s}\Phi(q_{i}-q_{s+1}),F_{s+1}(t)\big\}dx_{s+1},\quad s\geq1.
\end{eqnarray}
As above in case of the the Liouville equations \eqref{rivLiyv} the generating functional of solutions
of the BBGKY hierarchy \eqref{hierarBBGKI} satisfies equation \eqref{fh1}.

To clarify the structure of stated above evolution equations in functional derivatives we generalize
them for systems of particles interacting via many-body interaction potentials
\begin{eqnarray}\label{hamn}
   &&H_n=\sum_{i=1}^n K(p_i)+
      \sum_{k=1}^n\sum_{1=i_1<\ldots<i_k}^n\Phi^{(k)}(q_{i_1},\ldots,q_{i_k}),\quad n\geq1,
\end{eqnarray}
where $\Phi^{(k)}$ is a $k$-body interaction potential. In this case the Liouville equation in
functional derivatives has the form
\begin{eqnarray}\label{fh1nD}
   &&\frac{d}{dt}(D(t),u)=
      \int\big\{K(p_1),\frac{\delta}{\delta u(x_1)}(D(t),u)\big\}u(x_1)dx_1+\\
   &&+\sum_{n=1}^\infty\frac{1}{n!}\int\ldots\int\big\{\Phi^{(n)}(q_1,\ldots,q_n),
      \frac{\delta^n}{\delta u(x_1)\ldots\delta u(x_n)}(D(t),u)\big\}
      \prod_{i=1}^n u(x_i)dx_1\ldots dx_n,\nonumber
\end{eqnarray}
and distribution functions are determined by a sequence of the Liouville equations \eqref{rivLiyv}
with the Hamiltonian \eqref{hamn} respectively.

The BBGKY hierarchy in functional derivatives for the generating functional of marginal
distribution functions takes the form
\begin{eqnarray}\label{fh1nF}
   &&\hskip-8mm\frac{d}{dt}(F(t),u)=
     \int\big\{K(p_1),\frac{\delta}{\delta u(x_1)}(F(t),u)\big\}(u(x_1)+1)dx_1+\\
   &&\hskip-8mm+\sum_{n=1}^\infty\frac{1}{n!}\int\ldots\int\big\{\Phi^{(n)}(q_1,\ldots,q_n),
      \frac{\delta^n}{\delta u(x_1)\ldots\delta u(x_n)}(F(t),u)\big\}
      \prod_{i=1}^n (u(x_i)+1)dx_1\ldots dx_n,\nonumber
\end{eqnarray}
and the marginal distribution functions are determined by the BBGKY hierarchy respectively
\begin{eqnarray}\label{1b}
   &&\frac{\partial}{\partial t}F_{s}(t)=\big\{H_s,F_{s}(t)\big\}+
      \sum\limits_{k=1}^{s}\frac{1}{k!}\sum\limits_{i_1\neq\ldots\neq i_{k}=1}^{s}
      \,\sum\limits_{n=1}^{\infty}\frac{1}{n!}\int\ldots\int\big\{\Phi^{(k+n)}(q_{i_1},\ldots,q_{i_k},\\
   &&q_{s+1},\ldots,q_{s+n}),F_{s+n}(t)\big\}dx_{s+1}\ldots dx_{s+n},\quad s\geq1.\nonumber
\end{eqnarray}
We can see that the structure of the Liouville equation and the BBGKY hierarchy in functional derivatives
is similar in contrast to the structure of these equations for corresponding distribution functions.

\subsection{A nonperturbative solution of the BBGKY hierarchy}
On the basis of a solution of the Liouville equation  \eqref{rivLiyv} and relation (\ref{FuconDu})
for generating functionals, we construct a nonperturbative solution of the Cauchy problem of the
BBGKY hierarchy \eqref{hierarBBGKI}. We consider the sequence
\begin{eqnarray*}
  &&\widetilde{D}(t)=
     (S(-t)D(0),I)^{-1}(1,(S_1(-t)D_1(0))(x_1),\dots,(S_n(-t)D_n(0))(x_1,\ldots,x_n),\ldots),
\end{eqnarray*}
where $S_n(-t),\,t\in \mathbb{R}$, is the group of evolution operators defined on the space $L^1_n$
by the formula
\begin{eqnarray}\label{Sdef}
  &&(S_n(-t)f_n)(x_1,\ldots,x_n)\doteq f_n(X_1(-t,x_1,\ldots,x_n),\ldots,X_n(-t,x_1,\ldots,x_n)),
\end{eqnarray}
where $X_i(-t,x_1,\ldots,x_n),\, 1\leq i\leq n$, are solutions of the Cauchy problem of the Hamilton
equations of $n$ particle system with initial data: $x_1,\ldots,x_n$. The function $S_n(-t)D_n(0)$ is
a solution of the Cauchy problem of the Liouville equation \eqref{rivLiyv} with the initial data
$D_n(0)$ \cite{CGP97}.

In the functional $(e^\mathfrak{a}S(-t)D(0),u)$ we expand operators \eqref{Sdef} over their cumulants as
the following cluster expansions
\begin{eqnarray}\label{cexp}
   &&\hskip-5mm S_{s+n}(-t,Y,\,X\setminus Y)=\sum\limits_{\mathrm{P}:\,(\{Y\},\,X\setminus Y)=
       \bigcup_i X_i}\,\prod\limits_{X_i\subset\mathrm{P}}\mathfrak{A}_{|X_i|}(-t,X_i),\quad n\geq0,
\end{eqnarray}
where $Y\equiv(x_1,\ldots,x_s)$, $(\{Y\})$ is the set consisting of one element $Y=(x_1,\ldots,x_s)$,
$X\equiv(x_1,\ldots,x_{s+n})$, and ${\sum_{\mathrm{P}:(\{Y\},\,X\setminus Y)={\bigcup_i} X_i}}$ is the
sum over all possible partitions $\mathrm{P}$ of the set $(\{Y\},\,X\setminus Y)$ into $|\mathrm{P}|$
nonempty mutually disjoint subsets $X_i\subset(\{Y\},\,X\setminus Y)$. Owing to the equality
\begin{eqnarray*}
   &&\hskip-7mm\sum\limits_{\mathrm{P}:\,(\{Y\},\,X\setminus Y)=
       \bigcup_i X_i}\,\prod\limits_{X_i\subset\mathrm{P}}\mathfrak{A}_{|X_i|}(-t,X_i)=
       \sum_{Z\subset X\setminus Y}\mathfrak{A}_{1+|Z|}(-t,\{Y\},Z)\sum\limits_{\mathrm{P}:
       \,X\setminus Y\setminus Z=\bigcup_i X_i}\,\prod\limits_{X_i\subset\mathrm{P}}\mathfrak{A}_{|X_i|}(-t,X_i),
\end{eqnarray*}
and as a result of the symmetry property of the integrand the validity of the following equality
\begin{eqnarray*}
  &&\hskip-5mm\sum_{Z\subset X\setminus Y}\mathfrak{A}_{1+|Z|}(-t,\{Y\},Z)
       \sum\limits_{\mathrm{P}:\,X\setminus Y\setminus Z=\bigcup_i X_i}\,
       \prod\limits_{X_i\subset\mathrm{P}}\mathfrak{A}_{|X_i|}(-t,X_i)=\\
  &&\hskip-5mm=\sum\limits_{k=0}^{n}\sum\limits_{i_1<\ldots<i_k=1}^{n}
       \mathfrak{A}_{1+k}(-t,\{Y\},x_{s+1},\ldots,x_{s+k})
       \sum\limits_{\mathrm{P}:\,(x_{s+k+1},\ldots,x_{s+n})=\bigcup_i X_i}\,
       \prod\limits_{X_i\subset\mathrm{P}}\mathfrak{A}_{|X_i|}(-t,X_i),
\end{eqnarray*}
we have
\begin{eqnarray*}
  &&(e^\mathfrak{a}S(-t)D(0),u)=\sum_{s=0}^\infty\frac{1}{s!}
     \sum_{n=0}^\infty\frac{1}{n!}\int\ldots\int
     \mathfrak{A}_{1+|X\setminus Y|}(-t,\{Y\},\,X\setminus Y)\sum_{k=0}^\infty\frac{1}{k!}\times\\
  &&\times\sum\limits_{\mathrm{P}:\,(x_{s+n+1},\ldots,x_{s+n+k})=
       \bigcup_i X_i}\,\prod\limits_{X_i\subset\mathrm{P}}\mathfrak{A}_{|X_i|}(-t,X_i)D_{s+n+k}(0)
       \prod_{i=1}^s u(x_i)dx_{1}\ldots dx_{s+n+k}.
\end{eqnarray*}
Here the solution of cluster expansions \eqref{cexp}
\begin{eqnarray}\label{cum}
   &&\hskip-5mm\mathfrak{A}_{1+n}(-t,\{Y\},\,X\setminus Y)=\sum\limits_{\mathrm{P}:\,(\{Y\},\,X\setminus Y)=
       \bigcup_i X_i}(-1)^{|\mathrm{P}|-1}({|\mathrm{P}|-1})!
       \prod\limits_{X_i\subset\mathrm{P}}S_{|\theta(X_i)|}(-t,\theta(X_i))
\end{eqnarray}
is the $(1+n)th$-order cumulant of groups of operators \eqref{Sdef}, and we introduced the declasterization
mapping by the following formula: $\theta(\{Y\},X\setminus Y)=X$.

According to the validity of the equality
\begin{eqnarray*}
  &&\int\ldots\int\sum\limits_{\mathrm{P}:\,(x_{s+n+1},\ldots,x_{s+n+k})=
    \bigcup_i X_i}\,\prod\limits_{X_i\subset\mathrm{P}}\mathfrak{A}_{|X_i|}(-t,X_i)D_{s+n+k}(0)
    dx_{s+n+1}\ldots dx_{s+n+k}=\\
  &&=\int\ldots\int D_{s+n+k}(0)dx_{s+n+1}\ldots dx_{s+n+k},
\end{eqnarray*}
and the similar equality for the normalizing factor \cite{CGP97}
\begin{eqnarray*}
  &&(S(-t)D(0),I)=(D(0),I),
\end{eqnarray*}
and taking into account the definition of initial marginal distribution functions, from relation
\eqref{FuconDu} we derive
\begin{eqnarray}\label{mdfs}
  &&F_s(t,Y)=\sum_{n=0}^\infty\frac{1}{n!}\int\ldots\int
     \mathfrak{A}_{1+n}(-t,\{Y\},\,X\setminus Y)F_{s+n}(0,X)dx_{s+1}\ldots dx_{s+n},
\end{eqnarray}
where $\mathfrak{A}_{1+n}(-t,\{Y\},\,X\setminus Y)$ is $(1+n)th$-order cumulant \eqref{cum}.

In fact the following criterion is valid. Expansion (\ref{mdfs}) is a solution of the initial-value problem
of the BBGKY hierarchy \eqref{1b} if and only if the evolution operators $\mathfrak{A}_{1+n}(-t),\,n\geq0$,
satisfy the recurrence relations \eqref{cexp}.

We indicate that nonperturbative solution \eqref{mdfs} of the BBGKY hierarchy is transformed to the
form of the perturbation (iteration) series as a result of applying of analogs of the Duhamel equation
to cumulants \eqref{cum} of groups of operators \cite{GR,GerRS}.

For the first time a nonperturbative solution of the BBGKY hierarchy was constructed in paper \cite{L}
(see also \cite{CGP97}) in the form of an expansion over particle clusters which evolution is governed by
the corresponding-order reduced cumulant of groups of operators \eqref{Sdef}
\begin{eqnarray}\label{Udef}
  &&U_{1+n}(-t,\{Y\},X\setminus Y)=\sum^n_{k=0}(-1)^k \frac{n!}{k!(n-k)!}
     S_{s+n-k}(-t,Y,x_{s+1},\ldots,x_{s+n-k}).
\end{eqnarray}
Indeed, using the reduced cluster expansions of operators \eqref{Sdef} over reduced cumulants
\begin{eqnarray*}
  &&S_{s+n}(-t,Y,X\setminus Y)=
    \sum^n_{k=0}\frac{n!}{k!(n-k)!}U_{1+k}(-t,\{Y\},x_{s+1},\ldots,x_{s+k}),
\end{eqnarray*}
for the generating functional $(\widetilde{D}(t),u+1)$ we have as above
\begin{eqnarray*}
  &&(\widetilde{D}(t),u+1)=(D(0),I)^{-1}(S(-t)D(0),u+1)=\\
  &&=(D(0),I)^{-1}\sum_{s=0}^\infty\frac{1}{s!}
     \sum_{n=0}^\infty\frac{1}{n!}\int\ldots\int
     U_{1+n}(-t)\sum_{k=0}^\infty\frac{1}{k!}D_{s+n+k}(0)\prod_{i=1}^s u(x_i)dx_{1}\ldots dx_{s+n+k},
\end{eqnarray*}
and thus,
\begin{eqnarray*}\label{rL}
  &&F_s(t,Y)=\sum_{n=0}^\infty\frac{1}{n!}\int\ldots\int
     U_{1+n}(-t,\{Y\},X\setminus Y)F_{s+n}(0,X)dx_{s+1}\ldots dx_{s+n},
\end{eqnarray*}
where $U_{1+n}(-t)$ is the $(1+n)th$-order reduced cumulant \eqref{Udef}.


\section{Evolution equations in functional derivatives for correlations}

\subsection{The Liouville hierarchy for correlation functions}
In addition to an approach of the description of the evolution of states of many-particle systems
in terms of distribution functions one more an equivalent approach is given by means of correlation
functions which governed by the Liouville hierarchy.

We introduce the generating functional $(g(t),u)$ of correlation functions $g_s(t,x_1,\ldots,x_s),\,s\geq1$,
by means of generating functional \eqref{func_stan} as follows
\begin{eqnarray}\label{Dg}
   &&(D(t),u)=e^{(g(t),u)}.
\end{eqnarray}
Then taking into account the equality for $k\geq1$
\begin{eqnarray*}
   &&\hskip-5mm\frac{\delta^k}{\delta u(x_1)\ldots\delta u(x_k)}(D(t),u)=
     e^{(g,u)}\hskip-2mm\sum_{\mbox{\scriptsize $\begin{array}{c}\mathrm{P}:(x_1,\ldots,x_k)={\bigcup_j} X_j,
     \\X_j\equiv(x_{i_1},\ldots,x_{i_{|X_j|}})\end{array}$}}\prod_{X_j\subset\mathrm{P}}
     \frac{\delta^{|X_j|}}{\delta u(x_{i_1})\ldots\delta u(x_{i_{|X_j|}})}(g(t),u),
\end{eqnarray*}
according to the Liouville equation in functional derivatives \eqref{func_liouv}, the
evolution of states of many-particle systems in terms of generating functional \eqref{Dg}
of correlation functions is governed by the following evolution equation in functional derivatives
\begin{eqnarray}\label{eqfg_2}
   &&\hskip-8mm\frac{d}{dt}(g(t),u)=
     \int\big\{K(p_1),\frac{\delta}{\delta u(x_1)}(g(t),u)\big\}u(x_1)dx_1+\\
   &&\hskip-8mm+\frac{1}{2!}\int\int\big\{\Phi(q_1-q_2),
     \big(\frac{\delta^{2}}{\delta u(x_{1})\delta u(x_{2})}(g(t),u)+
     \prod_{j=1}^2\frac{\delta}{\delta u(x_{j})}(g(t),u)\big)\big\}u(x_1)u(x_2)dx_1dx_2,\nonumber
\end{eqnarray}
or in case of the Hamiltonian \eqref{hamn} this equation has the form
\begin{eqnarray}\label{eqfg}
   &&\hskip-8mm\frac{d}{dt}(g(t),u)=
     \int\big\{K(p_1),\frac{\delta}{\delta u(x_1)}(g(t),u)\big\}u(x_1)dx_1
     +\sum_{n=1}^\infty\frac{1}{n!}\int\ldots\int\big\{\Phi^{(n)}(q_1,\ldots,q_n),\\
   &&\hskip-8mm\sum_{\mbox{\scriptsize$\begin{array}{c}\mathrm{P}:(x_1,\ldots,x_n)={\bigcup_j} X_j,
     \\ X_j\equiv(x_{i_1},\ldots,x_{i_{|X_j|}})\end{array}$}}\prod_{X_j\subset\mathrm{P}}
     \frac{\delta^{|X_j|}(g(t),u)}{\delta u(x_{i_1})\ldots\delta u(x_{i_{|X_j|}})}\big\}
     \prod_{i=1}^n u(x_i)dx_1\ldots dx_n,\nonumber
\end{eqnarray}
where ${\sum_{\mathrm{P}:(x_1,\ldots,x_n)={\bigcup_j} X_j}}$ is the sum over all possible partitions
$\mathrm{P}$ of the set $(x_1,\ldots,x_n)$ into $|\mathrm{P}|$ nonempty mutually disjoint subsets
$X_j\equiv(x_{i_1},\ldots,x_{i_{|X_j|}})\subset(x_1,\ldots,x_n)$.

Taking into account equality (\ref{def_funder}), we observe that the correlation functions
$g_s(t),\,s\geq1$, are determined by the Liouville hierarchy \cite{GerShJ,Sh}
\begin{eqnarray}\label{Lh}
   &&\hskip-5mm\frac{\partial}{\partial t}g_{s}(t,Y)=\big\{H_s,g_{s}(t,Y)\big\}+\\
   &&\hskip-5mm+\sum\limits_{\mbox{\scriptsize $\begin{array}{c}\mathrm{P}:Y=
     \bigcup_{i}X_{i},\\|\mathrm{P}|>1\end{array}$}}\hskip-2mm
     \sum\limits_{\mbox{\scriptsize$\begin{array}{c}{Z_{1}\subset X_{1}},\\
     Z_{1}\neq\emptyset\end{array}$}}\ldots\sum\limits_{\mbox{\scriptsize $\begin{array}{c} {Z_{|\mathrm{P}|}
     \subset X_{|\mathrm{P}|}},\\Z_{|\mathrm{P}|}\neq\emptyset \end{array}$}}
     \big\{\Phi^{(\sum\limits_{r=1}^{|\mathrm{P}|}|Z_{{r}}|)}
     (Z_{{1}},\ldots,Z_{{|\mathrm{P}|}}),\prod_{X_{i}\subset \mathrm{P}}g_{|X_{i}|}(t,X_{i})\big\},\nonumber
\end{eqnarray}
where notations accepted above are used.

We note that the correlation functions $g_{n}(t),\,n\geq1$, are interpreted as the functions that describe
correlations of states $D_{n}(t), n\geq1,$ governed by the Liouville equations \eqref{rivLiyv}. Indeed,
according to definition \eqref{def_funder}, from relation \eqref{Dg} for generating functionals we obtain
the cluster expansions of distribution functions over the correlation functions
\begin{eqnarray}\label{clexp}
    &&D_{s}(t,Y)=\sum\limits_{\mbox{\scriptsize $\begin{array}{c}\mathrm{P}:Y=\bigcup_{i}X_{i}\end{array}$}}
        \prod_{X_i\subset \mathrm{P}}g_{|X_i|}(t,X_i), \quad s\geq1,
\end{eqnarray}
i.e. the correlation functions are the cumulants (semi-invariants) of distribution functions
\begin{eqnarray*}\label{gBigfromDFB}
    &&g_{s}(t,Y)=\sum\limits_{\mbox{\scriptsize $\begin{array}{c}\mathrm{P}:Y=\bigcup_{i}X_{i}\end{array}$}}
        (-1)^{|\mathrm{P}|-1}(|\mathrm{P}|-1)!\,\prod_{X_i\subset \mathrm{P}}D_{|X_i|}(t,X_i), \quad s\geq1.
\end{eqnarray*}
The validity of relations \eqref{clexp} is a consequence of the following equality
\begin{eqnarray}\label{gfg}
   &&e^{(g(t),u)}=({\mathbb E}\mathrm{xp}_{\ast}g(t),u),
\end{eqnarray}
where we denote by $({\mathbb E}\mathrm{xp}_{\ast}g(t),u)$ the generating functional of the sequence
of functions
\begin{eqnarray*}
    &&({\mathbb E}\mathrm{xp}_{\ast}g(t))_{s}(Y)=
        \sum\limits_{\mbox{\scriptsize $\begin{array}{c}\mathrm{P}:Y=\bigcup_{i}X_{i}\end{array}$}}
        \prod_{X_i\subset \mathrm{P}}g_{|X_i|}(t,X_i), \quad s\geq1.
\end{eqnarray*}
Indeed, on sequences of functions $f,\widetilde{f}\in L^{1}$ we define the $\ast$-product
\begin{eqnarray}\label{Product}
    &&(f\ast\widetilde{f})_{|Y|}(Y)=\sum\limits_{Z\subset Y}\,f_{|Z|}(Z)
        \,\widetilde{f}_{|Y\setminus Z|}(Y\setminus Z),
\end{eqnarray}
where $\sum_{Z\subset Y}$ is the sum over all subsets $Z$ of the set $Y\equiv(x_1,\ldots,x_s)$.
By means of definition \eqref{Product} of the $\ast$-product we introduce on sequences
$f=(0,f_1(x_1),\ldots,f_n(x_1,\ldots,x_n),\ldots)$ the mapping ${\mathbb E}\mathrm{xp}_{\ast}$
by the expansions
\begin{eqnarray}\label{circledExp}
   &&({\mathbb E}\mathrm{xp}_{\ast}\,f )_{|Y|}(Y)=\big(\mathbb{I}+
      \sum\limits_{n=1}^{\infty} \frac{1}{n!}f^{\ast n}\big)_{|Y|}(Y)=\\
   &&=1\delta_{|Y|,0}+\sum\limits_{\mathrm{P}:\,Y=\bigcup_{i}X_{i}}\,
      \prod_{X_i\subset \mathrm{P}}f_{|X_i|}(X_i),\nonumber
\end{eqnarray}
where we use the notations accepted above, $\delta_{|Y|,0}$ is a Kronecker symbol and
$\mathbb{I}\equiv(1,0,\ldots,0,\ldots)$. Then observing the validity of the equality
\begin{eqnarray}\label{efg}
    &&(f\ast \widetilde{f},u)=(f,u)(\widetilde{f},u),
\end{eqnarray}
in view of \eqref{circledExp} we justify equality \eqref{gfg}.

A solution of the Cauchy problem of the Liouville hierarchy (\ref{Lh}) for the initial data $g(0)$
is defined by the following evolution operator \cite{GerShJ}
\begin{eqnarray}\label{rozvNF-N_F}
    &&g_{s}(t,Y)=\mathcal{S}(t,Y|g(0))\equiv\sum\limits_{\mathrm{P}:\,Y=\bigcup_i X_i}
      \mathfrak{A}_{|\mathrm{P}|}(-t,\{X_1\},\ldots,\{X_{|\mathrm{P}|}\})
      \prod_{X_i\subset \mathrm{P}}g_{|X_i|}(0,X_i).
\end{eqnarray}
Here $\mathfrak{A}_{|\mathrm{P}|}(-t)$ is the $|\mathrm{P}|th$-order cumulant of groups of operators
\eqref{Sdef} given by the expansion
\begin{eqnarray} \label{cumulantP}
    &&\mathfrak{A}_{|\mathrm{P}|}(-t,\{X_1\},\ldots,\{X_{|\mathrm{P}|}\})\doteq\\
    &&\doteq\sum\limits_{\mathrm{P}^{'}:\,(\{X_1\},\ldots,\{X_{|\mathrm{P}|}\})=
       \bigcup_k Z_k}(-1)^{|\mathrm{P}^{'}|-1}({|\mathrm{P}^{'}|-1})!
       \prod\limits_{Z_k\subset\mathrm{P}^{'}}S_{|\theta(Z_{k})|}(-t,\theta(Z_{k})),\nonumber
\end{eqnarray}
where $\sum_{\mathrm{P}^{'}:\,(\{X_1\},\ldots,\{X_{|\mathrm{P}|}\})=\bigcup_k Z_k}$ is the sum over
all possible partitions $\mathrm{P}^{'}$ of the set $(\{X_1\},\ldots,$ $\{X_{|\mathrm{P}|}\})$ into
$|\mathrm{P}^{'}|$ nonempty mutually disjoint subsets $Z_k\subset(\{X_1\},\ldots,$ $\{X_{|\mathrm{P}|}\})$.

Really, taking into account formula \eqref{Sdef} for the solution of the Liouville equations and the
cluster expansions of distribution functions over correlation functions at initial time:
$D(0)={\mathbb E}\mathrm{xp}_{\ast}g(0)$, from equality \eqref{gfg} we obtain
\begin{eqnarray*}
   &&(g(t),u)=\big({\mathbb L}\mathrm{n}_{\ast}(\mathbb{I}+
      S(-t){\mathbb E}\mathrm{xp}_{\ast}g(0)),u\big),
\end{eqnarray*}
where the inverse mapping ${\mathbb L}\mathrm{n}_{\ast}$ to mapping \eqref{circledExp} is defined by
\begin{eqnarray*}\label{circledLn}
   &&({\mathbb L}\mathrm{n}_{\ast}\,(\mathbb{I}+f))_{|Y|}(Y)=\big(\sum\limits_{n=1}^{\infty}
      \frac{(-1)^{n-1}}{n}f^{\ast n}\big)_{|Y|}(Y)=\\
   &&=\sum\limits_{\mathrm{P}:\,Y=\bigcup_{i}X_{i}}\,(-1)^{|\mathrm{P}|-1}(|\mathrm{P}|-1)!
      \prod_{X_i\subset \mathrm{P}}f_{|X_i|}(X_i).\nonumber
\end{eqnarray*}
As a result of regrouping terms with respect to the same initial data we derive formula \eqref{rozvNF-N_F}
in componentwise form.

\subsection{Some relations between generating functionals}
Starting from an alternative approach to the description of states by the generating functional
of correlation functions \eqref{Dg}, we define the generating functional of marginal distribution
functions by means of such generating functional. Moreover we introduce also the generating functional
of the marginal correlation functions and consider relations of these generating functionals.

According to relations \eqref{FuconDu} and \eqref{Dg}, the generating functional of marginal distribution
functions is determined by means of generating functional of correlation functions as follows
\begin{eqnarray}\label{Fg}
   &&(F(t),u)=e^{(g(t),u+1)-(g(t),I)}.
\end{eqnarray}
Using this relation, the BBGKY hierarchy in functional derivatives \eqref{fh1nF} can be derived on the
basis of the Liouville hierarchy in functional derivatives \eqref{eqfg_2}.

From \eqref{Fg} in view of definition \eqref{def_funder} we obtain the expression for marginal distribution
functions in terms of correlation functions
\begin{eqnarray}\label{Fgn}
    &&F_{s}(t,Y)\doteq\sum\limits_{n=0}^{\infty}\frac{1}{n!}
      \int\ldots\int g_{1+n}(t,\{Y\},x_{s+1},\ldots,x_{s+n})dx_{s+1}\ldots dx_{s+n},\quad s\geq1,
\end{eqnarray}
where we denote by $(\{Y\})$ the set consisting of one element $Y=(1,\ldots,s)$, and the correlation
functions $g_{1+n}(t,\{Y\},x_{s+1},\ldots,x_{s+n}),\,n\geq0$, are solutions of the Liouville hierarchy
\eqref{Lh} in case of particle cluster of $s$ particles and particles \cite{GP}. To prove relation
(\ref{Fgn}) we transform the functional from the right-hand side of relation \eqref{Fg} to the canonical
form using equalities \eqref{gfg} and \eqref{r1}
\begin{eqnarray*}
    &&e^{(g(t),u+1)-(g(t),I)}=({\mathbb E}\mathrm{xp}_{\ast}g(t),I)^{-1}({\mathbb E}\mathrm{xp}_{\ast}g(t),u+1)=\\
    &&=({\mathbb E}\mathrm{xp}_{\ast}g(t),I)^{-1}(e^{\mathfrak{a}}{\mathbb E}\mathrm{xp}_{\ast}g(t),u).
\end{eqnarray*}
The last functional is the generating functional of the sequence of functions
\begin{eqnarray*}
    &&({\mathbb E}\mathrm{xp}_{\ast}g(t),I)^{-1}e^{\mathfrak{a}}{\mathbb E}\mathrm{xp}_{\ast}g(t).
\end{eqnarray*}
To write down it in the component-wise form on measurable functions we introduce the following mappings
\begin{eqnarray*}
    &&(\mathfrak{d}_{Y} f)_{n}\doteq f_{|Y|+n}(Y,x_{s+1},\ldots,x_{s+n}),\\
    &&(\mathfrak{d}_{\{Y\}} f)_{n}\doteq f_{1+n}(\{Y\},x_{s+1},\ldots,x_{s+n}),\quad n\geq0.
\end{eqnarray*}
Then we have
\begin{eqnarray*}
   &&({\mathbb E}\mathrm{xp}_{\ast}g(t),I)^{-1}(e^{\mathfrak{a}}{\mathbb E}\mathrm{xp}_{\ast}g(t))_{s}(Y)=
       ({\mathbb E}\mathrm{xp}_{\ast}g(t),I)^{-1}(\mathfrak{d}_{Y}{\mathbb E}\mathrm{xp}_{\ast}g(t),I).
\end{eqnarray*}
Owing the validity of the following equalities \cite{GP}
\begin{eqnarray*}
  &&\mathfrak{d}_{Y}{\mathbb E}\mathrm{xp}_{\ast}g(t)=
       \mathfrak{d}_{\{Y\}}{\mathbb E}\mathrm{xp}_{\ast}g(t),
\end{eqnarray*}
and
\begin{eqnarray*}
  &&\mathfrak{d}_{\{Y\}}{\mathbb E}\mathrm{xp}_{\ast}g(t)=
       {\mathbb E}\mathrm{xp}_{\ast}g(t)\ast\mathfrak{d}_{\{Y\}}g(t),
\end{eqnarray*}
and according to equality \eqref{efg}, we finally derive \eqref{Fgn}
\begin{eqnarray*}
  &&({\mathbb E}\mathrm{xp}_{\ast}g(t),I)^{-1}(\mathfrak{d}_{Y}{\mathbb E}\mathrm{xp}_{\ast}g(t),I)=
     (\mathfrak{d}_{\{Y\}}g(t),I).
\end{eqnarray*}

On the basis of solution \eqref{rozvNF-N_F} of the Liouville hierarchy \eqref{eqfg_2} and relation (\ref{Fg})
for generating functionals, we also derive nonperturbative solution \eqref{mdfs} of initial-value problem of
the BBGKY hierarchy \eqref{hierarBBGKI}.

By analogy with relation \eqref{FuconDu} we introduce the generating functional of the marginal correlation
functions by means of generating functional of correlation functions as follows
\begin{eqnarray}\label{Gg}
  &&(G(t),u)=(g(t),u+1)-(g(t),I),
\end{eqnarray}
or in view of definition \eqref{def_funder} in componentwise form it has the form
\begin{eqnarray}\label{Ggn}
   &&G_{s}(t,x_1,\ldots,x_s)\doteq\sum\limits_{n=0}^{\infty}\frac{1}{n!}
     \int\ldots\int g_{s+n}(t,x_1,\ldots,x_{s+n})dx_{s+1}\ldots dx_{s+n},\quad s\geq1,
\end{eqnarray}
where the correlation functions $g_{s+n}(t)$ are solutions of the Liouville hierarchy \eqref{Lh}. We
emphasize that every term of marginal correlation function expansion \eqref{Ggn} is determined by the
$(s+n)th$-particle correlation function as contrasted to marginal distribution function expansion
\eqref{Fgn} which is defined by the $(1+n)th$-particle correlation function.

From relations \eqref{Gg} and \eqref{Fg} we obtain the relation of the generating functionals of
marginal distribution  functions and marginal correlation functions
\begin{eqnarray}\label{FGgf}
   &&(F(t),u)=e^{(G(t),u)}.
\end{eqnarray}
We note that in view of definition \eqref{def_funder} in componentwise form relation \eqref{FGgf} is
the cluster expansions of the marginal distribution functions over the marginal correlation functions
(it is an analog of cluster expansions \eqref{clexp} or \eqref{Dg})
\begin{eqnarray*}\label{FG}
   &&F_{s}(t,Y)=\sum\limits_{\mbox{\scriptsize $\begin{array}{c}\mathrm{P}:Y=\bigcup_{i}X_{i}\end{array}$}}
      \prod_{X_i\subset \mathrm{P}}G_{|X_i|}(t,X_i),\quad s\geq1,
\end{eqnarray*}
where ${\sum\limits}_{\mathrm{P}:Y=\bigcup_{i} X_{i}}$ is the sum over all possible partitions $\mathrm{P}$
of the set $Y\equiv(x_1,\ldots,x_s)$ into $|\mathrm{P}|$ nonempty mutually disjoint subsets $X_i\subset Y$.

\subsection{The nonlinear BBGKY hierarchy in functional derivatives}
From relation \eqref{Gg} we derive the nonlinear BBGKY hierarchy in functional derivatives for the
generating functional of marginal correlation functions on the basis of the Liouville hierarchy in
functional derivatives \eqref{eqfg_2}
\begin{eqnarray}\label{eqfG_2}
   &&\hskip-8mm\frac{d}{dt}(G(t),u)=
     \int\big\{K(p_1),\frac{\delta}{\delta u(x_1)}\big\}(G(t),u)(u(x_1)+1)dx_1+\\
   &&\hskip-8mm+\frac{1}{2!}\int\int\big\{\Phi(q_1-q_2),
     \big(\frac{\delta^{2}}{\delta u(x_{1})\delta u(x_{2})}(G(t),u)+
     \prod_{j=1}^2\frac{\delta}{\delta u(x_{j})}(G(t),u)\big)\big\}
     \prod_{i=1}^2(u(x_i)+1)dx_1dx_2,\nonumber
\end{eqnarray}
or in case of the Hamiltonian \eqref{hamn} this equation has the form
\begin{eqnarray}\label{eqfG}
   &&\hskip-8mm\frac{d}{dt}(G(t),u)=
     \int\big\{K(p_1),\frac{\delta}{\delta u(x_1)}(G(t),u)\big\}(u(x_1)+1)dx_1+\\
   &&\hskip-8mm+\sum_{n=1}^\infty\frac{1}{n!}\int\ldots\int\big\{\Phi^{(n)}(q_1,\ldots,q_n),
     \sum_{\mbox{\scriptsize$\begin{array}{c}\mathrm{P}:(x_1,\ldots,x_n)=
     {\bigcup_j}X_j,\\X_j\equiv(x_{i_1},\ldots,x_{i_{|X_j|}})\end{array}$}}\prod_{X_j\subset\mathrm{P}}
     \frac{\delta^{|X_j|}(G(t),u)}{\delta u(x_{i_1})\ldots\delta u(x_{i_{|X_j|}})}\big\}\times \nonumber\\
   &&\hskip-8mm\times\prod_{i=1}^n(u(x_i)+1)dx_1\ldots dx_n,\nonumber
\end{eqnarray}
where we use notations accepted above in equation \eqref{eqfg}.

Taking into account equality (\ref{def_funder}), we observe that the marginal correlation functions
are determined by the nonlinear BBGKY hierarchy
\begin{eqnarray}\label{nhBBGKY}
   &&\hskip-5mm\frac{\partial}{\partial t} G_s(t,Y)=\big\{H_s,G_{s}(t,Y)\big\}+\\
   &&\hskip-5mm+\sum\limits_{\mbox{\scriptsize $\begin{array}{c}\mathrm{P}:Y=
     \bigcup_{i}X_{i},\\|\mathrm{P}|>1\end{array}$}}\hskip-2mm
     \sum\limits_{\mbox{\scriptsize$\begin{array}{c}{Z_{1}\subset X_{1}},\\
     Z_{1}\neq\emptyset\end{array}$}}\hskip-2mm\ldots\hskip-2mm
     \sum\limits_{\mbox{\scriptsize $\begin{array}{c} {Z_{|\mathrm{P}|}\subset X_{|\mathrm{P}|}},\\
     Z_{|\mathrm{P}|}\neq\emptyset \end{array}$}}
     \hskip-2mm \big\{\Phi^{(\sum\limits_{r=1}^{|\mathrm{P}|}|Z_{{r}}|)}
     (Z_{{1}},\ldots,Z_{{|\mathrm{P}|}}),\prod_{X_{i}\subset \mathrm{P}}G_{|X_{i}|}(t,X_{i})\big\}+\nonumber\\
   &&\hskip-5mm+\sum_{n=1}^\infty\sum_{k=1}^n\sum_{1=j_1<\ldots<j_k}^s\int\ldots\int
     \big\{\Phi^{(n+1)}(q_{j_1},\ldots,q_{j_k},q_{s+1},\ldots,q_{s+n+1-k}),\nonumber\\
   &&\hskip-5mm\sum_{\mbox{\scriptsize$\begin{array}{c}\mathrm{P}:(x_{1},\ldots,x_{s+n+1-k})=
     {\bigcup_{j}}X_j,|\mathrm{P}|\leq n+1,\\X_j\not\subseteq Y\setminus(x_{j_1},\ldots,x_{j_k})\end{array}$}}
     \prod_{X_j\subset\mathrm{P}}G_{|X_j|}(t,X_j)\big\}dx_{s+1}\ldots dx_{s+n+1-k}.\nonumber
\end{eqnarray}

We can see that the structure of the Liouville hierarchy \eqref{eqfg_2} and the nonlinear BBGKY
hierarchy \eqref{eqfG_2} in functional derivatives is similar in contrast to the structure of these
hierarchies for corresponding correlation functions.

On the basis of solution \eqref{rozvNF-N_F} of the Liouville hierarchy \eqref{Lh} and relation (\ref{Gg})
for generating functionals, as above in case of relation (\ref{FuconDu}) we derive a nonperturbative
solution of initial-value problem of the nonlinear BBGKY hierarchy \eqref{nhBBGKY}.

Taking into account the equality
\begin{eqnarray}\label{solG1}
   &&(g(t),u+1)=(\mathcal{S}(t|g(0)),u+1)=(e^{\mathfrak{a}}\mathcal{S}(t|g(0)),u),
\end{eqnarray}
in the functional $(e^{\mathfrak{a}}\mathcal{S}(t|g(0)),u)$ we expand operators \eqref{rozvNF-N_F} over
the corresponding cumulants. As a result from relation (\ref{Gg}) we derive a solution expansion of the
nonlinear BBGKY hierarchy \cite{GP11}
\begin{eqnarray}\label{sss}
    &&\hskip-8mmG_{s}(t,Y)=\sum\limits_{n=0}^{\infty}\frac{1}{n!}
        \,\int\ldots\int U_{1+n}(t;\{Y\},x_{s+1},\ldots,x_{s+n}\mid G(0))dx_{s+1}\ldots dx_{s+n},\quad s\geq1,
\end{eqnarray}
where the $(n+1)th$-order reduced cumulant $U_{1+n}(t)$ of groups \eqref{rozvNF-N_F} has been introduced
\begin{eqnarray}\label{ssss}
   &&\hskip-8mmU_{1+n}(t;\{Y\},x_{s+1},\ldots,x_{s+n} \mid G(0))\doteq \sum_{k=0}^n(-1)^k\frac{n!}{k!(n-k)!}\times\\
   &&\hskip-8mm\times\sum_{\mathrm{P}:(x_1,\ldots,x_{s+n-k})={\bigcup_i}
     X_i}\mathfrak{A}_{|\mathrm{P}|}(-t,X_1,\ldots,X_{|\mathrm{P}|})\sum_{k_1=0}^k\frac{k!}{k_1!(k-k_1)!}\ldots
     \sum_{k_{|\mathrm{P}|-1}=0}^k\frac{k_{|\mathrm{P}|-2}!}{k_{|\mathrm{P}|-1}!
     (k_{|\mathrm{P}|-2}-k_{|\mathrm{P}|-1})!}\nonumber\\
   &&\hskip-8mm\times G_{|X_1|+k-k_1}(0,X_1,x_{s+n-k+1},\ldots,x_{s+n-k_1})\ldots
     G_{|X_{|\mathrm{P}|}|+k_{|\mathrm{P}|-1}}(0,X_{|\mathrm{P}|},x_{s+n-k_{|\mathrm{P}|-1}+1},\ldots,x_{s+n}),\nonumber
\end{eqnarray}
and $\mathfrak{A}_{|\mathrm{P}|}(-t)$ is the $|\mathrm{P}|th$-order cumulant defined by the formula
\eqref{cumulantP}.

We give simplest examples of reduced nonlinear cumulants (\ref{ssss}):
\begin{eqnarray*}
    &&U_{1}(t;\{Y\}\mid G(0))=\mathcal{G}(t;Y\mid G(0))=\\
    &&=\sum\limits_{\mathrm{P}:\,Y=\bigcup_i X_i}
        \mathfrak{A}_{|\mathrm{P}|}(-t,\{X_1\},\ldots,\{X_{|\mathrm{P}|}\})
        \prod\limits_{X_i\subset\mathrm{P}}G_{|X_i|}(0,X_{i}),\\ \\
    &&U_{2}(t;\{Y\},x_{s+1}\mid G(0))=\sum\limits_{\mathrm{P}:\,(Y,x_{s+1})=\bigcup_i X_i}
        \mathfrak{A}_{|\mathrm{P}|}(-t,\{X_1\},\ldots,\{X_{|\mathrm{P}|}\})
        \prod\limits_{X_i\subset\mathrm{P}}G_{|X_i|}(0,X_{i})-\\
    &&-\sum\limits_{\mathrm{P}:Y=\bigcup_i X_i}\mathfrak{A}_{|\mathrm{P}|}(-t,\{X_1\},\ldots,
        \{X_{|\mathrm{P}|}\})\sum_{j=1}^{|\mathrm{P}|}
        \prod\limits_{\mbox{\scriptsize $\begin{array}{c}{X_i\subset\mathrm{P}},
        \\X_i\neq X_j\end{array}$}}G_{|X_i|}(0,X_{i})G_{|X_j|+1}(0,X_{j},x_{s+1}).
\end{eqnarray*}

In case of initial data satisfying a chaos property, i.e. $G^{(1)}(0)\equiv(0,G_{1}(0,1),0,\ldots)$,
for the $(1+n)th$-order reduced cumulant (\ref{ssss}) we have
\begin{eqnarray*}
  &&U_{1+n}(t;\{Y\},X\setminus Y \mid G^{(1)}(0))=
     \mathfrak{A}_{s+n}(-t,X)\prod\limits_{i=1}^{s+n}G_{1}(0,x_i),
\end{eqnarray*}
where $\mathfrak{A}_{s+n}(-t)$ is $(s+n)th$-order cumulant \eqref{cum}, and solution expansion \eqref{sss} takes
the more transparent structure. From the structure of this series it is clear that in case of absence of correlations
at initial time the correlations generated by the dynamics of many-particle systems are completely governed
by cumulants \eqref{cum} of groups of operators \eqref{Sdef}.


\section{The dual BBGKY hierarchy in functional derivatives}

\subsection{A generating functional of marginal observables}
The evolution of many-particle systems is described within the framework of the evolution of
states or in terms of an equivalent approach as the evolution of observables \cite{CGP97}.
The evolution of observables of a system of a non-fixed number of particles is described
by the sequences $A(t)=(A_0,A_1(t,x_1),\dots,A_n(t,x_1,\ldots,x_n),\ldots)$ of observables.
The bounded function $A_n(t,x_1,\ldots,x_n)$ is defined on the $n$-particle phase space and
symmetric with respect to the permutations of arguments $x_1,\ldots,x_s$. According to definition
(\ref{def_fun}) we introduce the generating functional $(A(t),u)$ of observables by the expansion
\begin{eqnarray}\label{func_spos}
  &&(A(t),u)\doteq\sum_{n=0}^\infty\frac{1}{n!}\int\ldots\int
     A_n(t,x_1,\ldots,x_n)\prod_{i=1}^nu(x_i)dx_1\ldots dx_n.
\end{eqnarray}
For the integrable functions $u(x_i),\,i\geq1$, functional \eqref{func_spos} exists.

Within the framework of generating functional (\ref{func_spos}) the evolution of observables
of many-particle systems with the Hamiltonian \eqref{ham1} is governed by the following evolution
equation in functional derivatives
\begin{eqnarray}\label{func_liouvdual}
  &&\frac{d}{d t}(A(t),u)=
    \int \big\{\frac{\delta}{\delta u(x_1)}(A(t),u),K(p_1)\big\}u(x_1)dx_1+\\
  &&+\frac{1}{2!}\int\int\big\{\frac{\delta^2}
    {\delta u(x_1)\delta u(x_2)}(A(t),u),\Phi(q_1-q_2)\big\}u(x_1)u(x_2)dx_1dx_2,\nonumber
\end{eqnarray}
where $\{\,.\,,\,.\,\}$ is the Poisson bracket. Taking into account equality (\ref{def_funder}),
we establish that observables $A_n(t),\,n\geq1$, are determined by the sequence of the Liouville
equations for observables
\begin{eqnarray}\label{rivLiyvdual}
   &&\frac{\partial}{\partial{t}}A_n(t)=\big\{A_n(t),H_n\big\}, \quad n\geq1,
\end{eqnarray}
or vice versa the generating functional of solutions of the Liouville equations \eqref{rivLiyvdual}
satisfies equation \eqref{func_liouvdual}.

We introduce a generating functional of marginal observables $B(t)=(B_0,B_1(t,x_1),\dots,$
$B_s(t,x_1,\ldots,x_s),\ldots)$ by means of generating functional \eqref{func_spos} as follows
\begin{eqnarray}\label{ffd}
  &&(B(t),u)=(A(t),u e^{-\int u(x)dx}),
\end{eqnarray}
where we use the notation
\begin{eqnarray*}
  &&u e^{-\int u(x)dx}=(1,u(x_1),\ldots,\prod_{i=1}^nu(x_i),\ldots)
    \sum_{k=0}^\infty\frac{(-1)^k}{k!}\int\ldots\int\prod_{j=1}^k u(x_j)dx_1\ldots dx_k.
\end{eqnarray*}
A generating functional of marginal observables gives an equivalent approach to the description
of the evolution of finitely many particles in addition to generating functional of observables
\eqref{func_spos} and it is adopted for the description of infinite-particle systems.

In consequence of introduced relation \eqref{ffd} we establish the relation between marginal observables
and observables. We transform generating functional $(A(t),u e^{-\int u(x)dx})$ to form \eqref{def_fun},
and thus, we determine functions generated by this functional. The following equality holds (the dual
analog of equality \eqref{r1})
\begin{eqnarray}\label{r2}
  &&(A(t),u e^{-\int u(x)dx})=(e^{-\mathfrak{a^{+}}}A(t),u),
\end{eqnarray}
where on $b\in L^{\infty}$ the operator $\mathfrak{a^{+}}$ is defined by the formula (an analog of the
creation operator \cite{CGP97}, which is an adjoint operator to operator \eqref{opann} in the sense of
mean value functional \cite{CGP97})
\begin{eqnarray*}\label{opercr}
   &&(\mathfrak{a}^{+}b)_{s}(x_{1},\ldots,x_{s})\doteq\sum_{j=1}^s\,b_{s-1}((x_{1},\ldots,x_{s})\setminus (x_j)).
\end{eqnarray*}
Indeed we have
\begin{eqnarray*}
  &&(A(t),u e^{-\int u(x)dx})=\sum_{n=0}^\infty \frac{1}{n!}\sum_{k=0}^\infty\frac{(-1)^k}{k!}\int\ldots\int
    A_n(t,x_1,\ldots,x_n)\prod_{i=1}^{k+n}u(x_i)dx_1\ldots dx_{k+n}=\\
  &&=\sum_{n=0}^\infty\frac{1}{n!}\int\ldots\int\sum_{k=0}^n(-1)^k\frac{n!}{k!(n-k)!}A_{n-k}(t,x_1,\ldots,x_{n-k})
    \prod_{i=1}^n u(x_i)dx_1\ldots dx_n,\nonumber
\end{eqnarray*}
and therefore in view of symmetry property of observables, according to definition \eqref{def_funder},
from relation \eqref{ffd} we obtain the expression for marginal ($s$-particle) observables within the
framework of nonequilibrium grand canonical ensemble \cite{BGer}
\begin{eqnarray}\label{moo}
 &&B_s(t,x_1,\ldots,x_s)=\sum_{n=0}^s\,\frac{(-1)^n}{n!}\sum_{j_1\neq\ldots\neq j_{n}=1}^s
        A_{s-n}\big(t,(x_1,\ldots,x_s)\setminus (x_{j_1},\ldots,x_{j_{n}})).
\end{eqnarray}
Using relation \eqref{ffd}, from the Liouville equation in functional derivatives for observables
\eqref{func_liouvdual}, we derive the evolution equation for the generating functional of marginal
observables
\begin{eqnarray}\label{fh2}
  &&\hskip-5mm\frac{d}{d t}(B(t),u)=\int\big\{\frac{\delta}{\delta u(x_1)}(B(t),u),K(p_1)\big\}u(x_1)dx_1+\\
  &&\hskip-5mm+\frac{1}{2!}\int\int\big\{\frac{\delta^2}{\delta u(x_1)\delta u(x_2)}(B(t),u)+
     \sum_{j=1}^2\frac{\delta}{\delta u(x_j)}(B(t),u),\Phi(q_1-q_2)\big\}\prod_{i=1}^2u(x_i)dx_1dx_2.\nonumber
\end{eqnarray}
We are referred to this evolution equation as the dual BBGKY hierarchy in functional derivatives.
Taking into account equality \eqref{def_funder}, from evolution equation \eqref{fh2} we establish
that the marginal observables $B_s(t),\,s\geq1$, are determined by the dual BBGKY hierarchy \cite{BGer,BG}
\begin{eqnarray}\label{dh}
   &&\frac{\partial}{\partial t}B_s(t,Y)=\big\{B_s(t,Y),H_s\big\}+
     \sum_{i\neq j=1}^s\big\{B_{s-1}(t,Y\setminus(x_i)),\Phi(q_i-q_j)\big\},\quad s\geq1,
\end{eqnarray}
where $Y\equiv(x_1,\ldots,x_s)$ and $Y\setminus(x_i)\equiv(x_1\ldots,x_{i-1},x_{i+1},\ldots,x_{s})$.
As above in case of the Liouville equations \eqref{rivLiyvdual} the generating functional of solutions
of the dual BBGKY hierarchy \eqref{dh} satisfies equation \eqref{fh2}.

To clarify the structure of stated above evolution equations in functional derivatives we generalize them
for a system of particles interacting via many-body interaction potentials. In this case the Liouville
equation in functional derivatives has the form
\begin{eqnarray}\label{fh1nA}
   &&\hskip-5mm\frac{d}{d t}(A(t),u)=
      \int\big\{\frac{\delta}{\delta u(x_1)}(A(t),u),K(p_1)\big\}u(x_1)dx_1+\\
   &&\hskip-5mm+\sum_{n=1}^\infty\frac{1}{n!}\int\ldots\int\big\{
      \frac{\delta^n}{\delta u(x_1)\ldots\delta u(x_n)}(A(t),u),\Phi^{(n)}(q_1,\ldots,q_n)\big\}
      \prod_{i=1}^n u(x_i)dx_1\ldots dx_n,\nonumber
\end{eqnarray}
and observables are determined by a sequence of the Liouville equations \eqref{rivLiyvdual} with
the Hamiltonian \eqref{hamn} respectively. The dual BBGKY hierarchy in functional derivatives for
the generating functional of marginal observables takes the form
\begin{eqnarray}\label{fh2g}
  &&\hskip-9mm\frac{d}{d t}(B(t),u)=\int\{\frac{\delta}{\delta u(x_1)}(B(t),u),K(p_1)\}u(x_1)dx_1+
    \sum_{n=1}^\infty\frac{1}{n!}\times\\
  &&\hskip-9mm\times\int\ldots\int\{\sum_{k=1}^n\sum_{i_1<\ldots<i_k=1}^n
    \frac{\delta^k}{\delta u(x_{i_1})\ldots\delta u(x_{i_k})}(B(t),u),
    \Phi^{(n)}(q_1,\ldots,q_n)\}\prod_{i=1}^n u(x_i)dx_1\ldots dx_n,\nonumber
\end{eqnarray}
and the marginal observables are governed by the dual BBGKY hierarchy \cite{BG}
\begin{eqnarray}\label{hBBGKId}
   &&\frac{\partial}{\partial t}B_{s}(t,Y)=\big\{B_{s}(t,Y),H_s\big\}+\\
   &&+\sum\limits_{n=1}^{s}\frac{1}{n!}
     \sum\limits_{k=n+1}^s \frac{1}{(k-n)!}\sum_{j_1\neq\ldots\neq j_{k}=1}^s
     \big\{B_{s-n}(t,Y\setminus (x_{j_{1}},\ldots,x_{j_{n}})),
     \Phi^{(k)}(x_{j_1},\ldots,x_{j_{k}})\big\}.\nonumber
\end{eqnarray}
As is obvious the evolution equations in functional derivatives for dual generating functionals
\eqref{func_stan},\eqref{func_spos} have the form \eqref{fh1nD},\eqref{fh1nA} and correspondingly for
dual generating functionals \eqref{FuconDu},\eqref{ffd} the hierarchies in functional derivatives have
the form \eqref{fh1nF},\eqref{fh2g}.

\subsection{A nonperturbative solution of the dual BBGKY hierarchy}
On the basis of a solution of the Liouville equation  \eqref{rivLiyvdual} and relation (\ref{ffd})
for generating functionals, we construct a nonperturbative solution of the Cauchy problem of the
dual BBGKY hierarchy \eqref{hBBGKId}. With this aim we consider the sequence
\begin{eqnarray*}
  &&A(t)=(A_0,S_1(t)A_1(0,x_1),\dots,S_n(t)A_n(0,x_1,\ldots,x_n),\ldots),
\end{eqnarray*}
where the evolution operator $S_n(t)$ is defined on the space $L^{\infty}_n$ by formula \eqref{Sdef},
and it is determined a solution of the Cauchy problem of the Liouville equation for observables of
the $n$-particle system with the initial data $A_n(0)$.
Using equality \eqref{r2}, in the functional $(e^{-\mathfrak{a}^{+}}S(t)A(0),u)$ we expand operators
$S_n(t),\,n\geq1$, over their cumulants as the following cluster expansions
\begin{eqnarray}\label{cexpd}
   &&(-1)^nS_{s-n}(t,Y\setminus X)=\\
   &&=\sum_{k=0}^n(-1)^{n-k}\sum_{i_1<\ldots<i_k\in(j_1,\ldots,j_{n})}
      \mathfrak{A}_{1+k}(t,\{Y\setminus(x_{i_1},\ldots,x_{i_k})\},x_{i_1},\ldots,x_{i_k}),\nonumber
\end{eqnarray}
where the abridged notations are introduced: $X\equiv(x_{j_1},\ldots,x_{j_{n}})\subset Y$,
$(x_{i_1},\ldots,x_{i_k})\subset X$. Then the following equalities are true
\begin{eqnarray*}
  &&\hskip-5mm(e^{-\mathfrak{a}^{+}}S(t)A(0),u)=\\
  &&\hskip-5mm=\sum_{s=0}^\infty\frac{1}{s!}
     \int\ldots\int\sum_{n=0}^s \frac{(-1)^{n}}{n!}
     \sum_{j_1\neq\ldots\neq j_n=1}^{s}S_{s-n}(t)A_{s-n}(0)(Y\setminus X)
     \prod_{i=1}^su(x_i)dx_{1}\ldots dx_s=\\
  &&\hskip-5mm=\sum_{s=0}^\infty\frac{1}{s!}\int\ldots\int\sum_{n=0}^s\,\frac{1}{n!}
      \sum_{j_1\neq\ldots\neq j_{n}=1}^s
      \mathfrak{A}_{1+n}(t,\{Y\setminus (x_{j_1},\ldots,x_{j_{n}})\},x_{j_1},\ldots,x_{j_{n}})\times\\
  &&\hskip-5mm\times\sum\limits_{k=0}^{s-n}\frac{(-1)^k}{k!}
     \sum\limits_{\mbox{\scriptsize$\begin{array}{c}{i_1\neq\ldots\neq i_{k}=1},
     \\i_1,\ldots,i_{k}\neq j_1,\ldots,j_{n}\end{array}$}}^s\,
     A_{s-n-k}(0,Y\setminus X\setminus (x_{i_1},\ldots,x_{i_{k}}))\prod_{i=1}^su(x_i)dx_{1}\ldots dx_s.
\end{eqnarray*}
Owing to the definition of initial marginal observables \eqref{moo}, from relation \eqref{ffd}
and equality \eqref{r2} we finally derive
\begin{eqnarray}\label{hdsol}
   &&B_{s}(t,Y)=\sum_{n=0}^s\,\frac{1}{n!}\sum_{j_1\neq\ldots\neq j_{n}=1}^s
      \mathfrak{A}_{1+n}(t,\{Y\setminus X\},X)\,B_{s-n}(0,Y\setminus X),
\end{eqnarray}
where $X\equiv(x_{j_1},\ldots,x_{j_{n}})\subset Y$ and the $(1+n)th$-order cumulant of groups of operators
\eqref{Sdef} is defined as follows \cite{GerRS},\cite{BG}
\begin{eqnarray}\label{cumulantd}
    &&\hskip-7mm\mathfrak{A}_{1+n}(t,\{Y\setminus X\},X)\doteq
       \sum\limits_{\mathrm{P}:\,(\{Y\setminus X\},\,X)={\bigcup}_i X_i}
       (-1)^{\mathrm{|P|}-1}({\mathrm{|P|}-1})!\prod_{X_i\subset \mathrm{P}}S_{|\theta(X_i)|}(t,\theta(X_i)),
\end{eqnarray}
and we use notations accepted above, for example,
\begin{eqnarray*}
    &&\mathfrak{A}_{1}(t,\{Y\})=S_{s}(t,Y),\\
    &&\mathfrak{A}_{2}(t,\{Y\setminus (x_j)\},x_j)=S_{s}(t,Y)-S_{s-1}(t,Y\setminus(x_j))S_{1}(t,x_j).
\end{eqnarray*}

To prove that cumulants \eqref{cumulantd} are solutions of recurrence relations \eqref{cexpd} we observe
that recurrence relations \eqref{cexpd} hold on functions $b_{s-n}(Y\setminus X)$, and thus, the following
identity is true
\begin{eqnarray*}
   &&(-1)^nS_{s-n}(t,Y\setminus X)\sum\limits_{\mathrm{P}:\,X=\bigcup_i
   X_i}(-1)^{\mathrm{|P|}}\mathrm{|P|}!\prod\limits_{X_i\subset\mathrm{P}}S_{|X_i|}(t,X_i)b_{s-n}(Y\setminus X)=\\
   &&=\sum_{k=0}^n(-1)^{n-k}\sum_{i_1<\ldots<i_k\in(j_1,\ldots,j_{n})}
      \mathfrak{A}_{1+k}(t,\{Y\setminus(x_{i_1},\ldots,x_{i_k})\},x_{i_1},\ldots,x_{i_k})b_{s-n}(Y\setminus X),
\end{eqnarray*}
where ${\sum_{\mathrm{P}:X={\bigcup_i} X_i}}$ is the sum over all possible partitions $\mathrm{P}$ of
the set $X$ into $|\mathrm{P}|$ nonempty mutually disjoint subsets $X_i\subset X$. These recurrence
relations are rearranged form of the cluster expansions of the evolution operators $S_s(t),\,s\geq1,$
\begin{eqnarray}\label{ced}
   &&S_{s}(t,Y\setminus X,\,X)=\sum\limits_{\mathrm{P}:\,(\{Y\setminus X\},\,X)=
       \bigcup_i X_i}\,\prod\limits_{X_i\subset\mathrm{P}}\mathfrak{A}_{|X_i|}(t,X_i),\quad |X|\geq0,
\end{eqnarray}
and hence we obtain expression \eqref{cumulantd} as a solution of recurrence relations \eqref{ced}.

In fact as above the following criterion is valid. Expansion (\ref{hdsol}) is a solution of the
initial-value problem of the dual BBGKY hierarchy \eqref{hBBGKId} if and only if the evolution
operators $\mathfrak{A}_{1+n}(t),\,n\geq0$, satisfy the recurrence relations \eqref{ced}.

We remark that an analog of expansion \eqref{mdfs}-\eqref{Udef} in case of a solution of the dual BBGKY
hierarchy \eqref{dh} is given by expansion \eqref{hdsol} with the following reduced cumulants of groups
of operators \eqref{Sdef}
\begin{eqnarray}\label{rcd}
  &&U_{1+n}(t,\{x_1,\dots,x_{s-n}\},x_{s-n+1},\dots,x_{s})=\sum^n_{k=0}(-1)^{k}
      \frac{n!}{k!(n-k)!}S_{s-k}(t).
\end{eqnarray}
In fact it is a consequence of the equivalence of corresponding generating functionals. Indeed,
solving recurrence relations \eqref{ced} with respect to the first-order cumulants for the separation
terms which are independent from the variables $Y\setminus X$
\begin{eqnarray*}
   &&\mathfrak{A}_{1+n}(t,\{Y\setminus X\},X)=
     \sum\limits_{\substack{Z\subset X}} \mathfrak{A}_{1}(t,\{Y\setminus X\cup Z\})
     \sum\limits_{\mathrm{P}:\,X \setminus Z ={\bigcup\limits}_i X_i}
     (-1)^{|\mathrm{P}|}\,|\mathrm{P}|!\,\prod_{i=1}^{|\mathrm{P}|}\mathfrak{A}_{1}(t,\{X_{i}\}),
\end{eqnarray*}
where ${\sum\limits}_{\substack{Z\subset X}}$ is the sum over all possible subsets $Z\subset X$ of
the set $X$, and, taking into account the identity
\begin{eqnarray*}
  &&\sum\limits_{\mathrm{P}:\,X\setminus Z ={\bigcup\limits}_i X_i}
     (-1)^{|\mathrm{P}|}\,|\mathrm{P}|!\,\prod_{i=1}^{|\mathrm{P}|}
     \mathfrak{A}_{1}(t,\{X_{i}\})b_{s-n}(Y\setminus X)=
     \sum\limits_{\mathrm{P}:\,X \setminus Z ={\bigcup\limits}_i X_i}
     (-1)^{|\mathrm{P}|}\,|\mathrm{P}|!\,b_{s-n}(Y\setminus X),
\end{eqnarray*}
and the equality
\begin{eqnarray*}
  &&\sum\limits_{\mathrm{P}:\,X \setminus Z =
      {\bigcup\limits}_i X_i}(-1)^{|\mathrm{P}|}\,|\mathrm{P}|!=(-1)^{|X \setminus Z|},
\end{eqnarray*}
we derive expansion \eqref{hdsol} over reduced cumulants \eqref{rcd} of the dual BBGKY hierarchy \eqref{dh}.

We note that for certain classes of observables solution expansion \eqref{hdsol} has more simple structure.
As we can see from \eqref{moo} one component sequences of marginal observables correspond to observables of
certain structure, namely the marginal observable $B^{(1)}=(0,a_{1}(x_1),0,\ldots)$ corresponds to the
additive-type observable $A^{(1)}=(0,a_{1}(x_1),\ldots,{\sum\limits}_{i=1}^{n}a_{1}(x_i),\ldots)$, and in
the general case the $k$-ary-type marginal observable $B^{(k)}=(0,\ldots,0,a_{k}(x_1,\ldots,x_k),0,\ldots)$
corresponds to the $k$-ary-type observable $A^{(k)}=(0,\ldots,0,a_{k}(x_1,\ldots,x_k),\ldots,\sum_{i_{1}<\ldots<i_{k}=1}^{n}a_s(x_{i_{1}},\ldots,x_{i_{k})},\ldots)$.
Then in case of additive-type observables solution expansion \eqref{hdsol} attains the form
\begin{eqnarray*}
  &&B^{(1)}_s(t,Y)=\mathfrak{A}_{s}(t,Y)\sum\limits_{i=1}^{s}a_{1}(x_i),
\end{eqnarray*}
where $\mathfrak{A}_{s}(t,Y)$ is the $sth$-order cumulant \eqref{cumulantd} of groups of operators \eqref{Sdef}.


\section{Conclusion}

By means of established relations for generating functionals \eqref{FuconDu},\eqref{Dg},\eqref{Gg}
and \eqref{ffd} the hierarchies of evolution equations in functional derivatives for generating
functionals of marginal states and observables of classical many-particle systems, namely the BBGKY
hierarchy for marginal distribution functions \eqref{fh1}, the Liouville hierarchy \eqref{eqfg_2} for
correlation functions, the nonlinear BBGKY hierarchy for marginal correlation functions \eqref{eqfG_2}
and the dual BBGKY hierarchy for marginal observables \eqref{fh2g}, were derived from the Liouville
equations in functional derivatives for generating functionals of distribution functions \eqref{func_liouv}
and observables \eqref{fh1nA} respectively. To clarify the structure of these hierarchies in functional
derivatives we generalized them for particles interacting via many body interaction potentials, in
particular we have equations \eqref{fh1nD},\eqref{fh1nF} for states, equations \eqref{eqfg},\eqref{eqfG}
for correlations of states and equations \eqref{fh1nA},\eqref{fh2g} for observables.

On the basis of cluster expansions \eqref{cexp} and \eqref{cexpd} of groups of operators which describe
dynamics of finitely many particles the nonperturbative solutions of the hierarchies of evolution
equations for states and observables were constructed. They are represented in the form of expansions
\eqref{mdfs} and \eqref{hdsol} over particle clusters which evolution is governed by the corresponding
order cumulant \eqref{cum} of groups of operators \eqref{Sdef} of the Liouville equations \eqref{rivLiyv}
in case of the BBGKY hierarchy \eqref{1b} and cumulant \eqref{cumulantd} of groups of operators of the
Liouville equations for observables \eqref{rivLiyvdual} in case of the dual BBGKY hierarchy \eqref{hBBGKId}
respectively.

Moreover, along with the definition within the framework of the non-equilibrium grand canonical ensemble
the marginal distribution functions can be defined in terms of dynamics of correlations \eqref{Fg} that
allows, in particular to give the rigorous meaning of the states for more general classes of functions than
the integrable functions. Using relation \eqref{Gg} of the generating functional of marginal correlation
functions and the generating functional of correlation functions governed by the Liouville hierarchy
\eqref{Lh}, a nonperturbative solution of the nonlinear BBGKY hierarchy for marginal correlation functions
\eqref{nhBBGKY} was constructed in the form of expansions \eqref{sss} over particle clusters which evolution
is governed by the corresponding order cumulant \eqref{ssss} of groups of operators \eqref{rozvNF-N_F} of
the Liouville hierarchy \eqref{Lh}.

Thus, the concept of cumulants of groups of operators forms the basis of established nonperturbative
solution expansions of hierarchies of evolution equations of many-particle systems.


\addcontentsline{toc}{section}{References}
\renewcommand{\refname}{References}

\end{document}